\documentclass[twocolumn,aps,amsmath,amssymb,showpacs]{revtex4}

\usepackage{graphicx}% Include figure files
\usepackage{dcolumn}% Align table columns on decimal point
\usepackage{bm}% bold math
\begin{document}

\title{Non-Markovian master equation for a damped oscillator with time-varying parameters}
\author{Kwong Wa Chang and C. K. Law}

\affiliation{Department of Physics and Institute of Theoretical
Physics, The Chinese University of Hong Kong, Shatin, Hong Kong SAR,
China}

\date{\today}
\begin{abstract}
We derive an exact non-Markovian master equation that generalizes
the previous work [Hu, Paz and Zhang, Phys. Rev. D {\bf 45}, 2843
(1992)] to damped harmonic oscillators with time-varying parameters.
This is achieved by exploiting the linearity of the system and
operator solution in Heisenberg picture. Our equation governs the
non-Markovian quantum dynamics when the system is modulated by
external devices. As an application, we apply our equation to parity
kick decoupling problems. The time-dependent dissipative
coefficients in the master equation are shown to be modified
drastically when the system is driven by $\pi$ pulses. For coherence
protection to be effective, our numerical results indicate that
kicking period should be shorter than memory time of the bath. The
effects of using soft pulses in an ohmic bath are also discussed.
\end{abstract}

\pacs{03.65.Yz, 42.50.Lc}

\maketitle

\section{Introduction}

An oscillator linearly coupled with a harmonic oscillator bath has
been an important model for studying quantum dissipation and
decoherence. While the solution for the damped harmonic oscillator
is well-known under Born- and/or Markov-approximation \cite{Breuer,
Weiss, Gardiner, Carmichael}, general treatments waiving those
approximations did not appear until studies of Wigner functions by
Haake and Reibold \cite{Haake} who addressed the issues in low
temperature and strong damping regimes. Exact non-Markovian master
equation for the model was later derived by Hu, Paz and Zhang (HPZ)
\cite{HPZ} who employed the path-integral approach for initially
factorizable states. The use of Wigner function \cite{Yu, Connell}
and characteristic functions \cite{Pereverzev} presented alternative
means to derive the master equation. The study was generalized by
Karrlein and Grabert \cite{Karrlein} with path-integral approach to
cover non-factorizable initial states, who also pointed out that the
exact Liouville operator for a damped harmonic oscillator is not
independent of initial system states in general. Recently,
non-Markovian master equations have been generalized to
two-oscillator problems in order to study quantum disentanglement
processes \cite{Hu2008, Oh2, Paris1, Paris2, Roncaglia}.

In parallel with these advances, there is a growing interest in
coherence protection due to the need of preserving quantum
information stored in a system that is inevitably coupled with its
environment \cite{Viola}. In particular, dynamical decoupling
techniques for two-level systems based on various pulses sequences
have been developed and optimized \cite{Agarwal, Uhrig1, Uhrig2,
Uhrig3, Kurizki}. For damped harmonic oscillator systems, parity
kicks by a stream of $\pi$ pulses is a conceptually simple strategy
in realizing dynamical decoupling \cite{Viola, Tombesi1}. Such a
strategy has been studied primarily in the $\delta$-pulse regime
\cite{Tombesi1, Tombesi2,ShiokawaHu}, but little was explored in the
soft-pulse regime. In order to address the problem one needs to
establish equations describing the dissipation dynamics for systems
driven by time-varying external fields.

The main purpose of this paper is to present an exact non-Markovian
master equation for an oscillator with time-varying parameters in
general. Such time-varying parameters correspond to the modulation
of oscillator frequency and parametric interactions strengths caused
by external devices. As a demonstrative example, we apply our master
equation to study the dynamics in `parity kick' problems. Numerical
results of dissipative coefficients in the master equation are found
to be drastically modified by the $\pi$ pulses. By solving the
master equation, we also address the effectiveness of coherence
protection, quantified by fidelity ${\cal F} (t)$ of system state,
at various kicking frequencies with soft pulses.

\section{The Master Equation}

\label{sec:model}

We consider an oscillator coupled linearly with an oscillator bath
the total Hamiltonian $H = H_{S} (t) + H_{B} + H_{I}$, where
\begin{eqnarray}
H_S (t)& = &  \frac{\lambda (t) }{2}a^{2}  + \frac{{\lambda
^{\ast} (t) }}{2}{a^{\dagger 2}}  + \left[ \omega_{0} + \Omega (t)  \right]
 a^{\dagger}a {\label{eq:HS}}\\
H_B & = & \sum_{k} \omega_{k} b_{k}^{\dagger} b_{k} \\
H_I & = & \sum_{k} g_{k} (a + a^{\dagger} ) (b_{k}+b_{k}^{\dagger}).
\end{eqnarray}
Here the system oscillator has a time-dependent natural frequency
$\omega_{0} + \Omega (t)$ and a parametric coupling strength
$\lambda (t)$. With these time-dependent parameters, we treat in
generality the time-dependent damped harmonic oscillator problem,
which distinguishes our work from previous studies. Physically, the
frequency shift $\Omega (t)$ can be induced by off-resonance driving
fields, and  $\lambda (t)$ describes time-dependent parametric
effects, for example in down conversion processes in nonlinear
optics. In our model the bath consists of a large number of
oscillators, with $\omega_{k}$ and real coupling strength $g_{k}$
being the frequency and coupling strength of the $k$-th mode
oscillator. We also define the annihilation operators of the system
oscillator and the bath oscillators by $a \equiv \sqrt{\frac{m
\omega_{0}}{2}} \left( X + \frac{iP}{m \omega_{0}} \right)$ and
$b_{k} \equiv \sqrt{\frac{m_{k} \omega_{k}}{2}} \left( q_{k} +
\frac{i p_{k}}{m_{k} \omega_{k}} \right)$. Here $X$ and $P$ are
position and momentum operators of system oscillator with mass $m$
and frequency $\omega_{0}$, while $q_{k}$ and $p_{k}$ are position
and momentum operators of $k$-th mode of bath oscillator with mass
$m_{k}$.

The bath structure is characterized by spectral density $J(\omega) \equiv \sum_{k} g_{k}^{2}
\delta (\omega - \omega_{k})$. In the limit $\omega_{k}$
becomes continuum, we adopt the commonly used spectral density of the form
\begin{equation}
J(\omega) = \eta \omega \left( \frac{\omega}{\omega_{c}} \right)^{n-1}
e^{-\frac{\omega}{\omega_{c}}}
\label{eq:spectraldensity}
\end{equation}
where $\eta$ is a dimensionless real number governing the strength
of system-bath coupling, and $\omega_{c}$ is the cut-off frequency.
Necessity and justification to introduce cut-off frequency have been
discussed in \cite{Weiss}. The exponent $n$ is a real number that
determines the $\omega$-dependence of $J(\omega)$ in the low
frequency region, and for physical baths $n \geq 0$. In literatures,
$0<n<1$, $n=1$ and $n>1$ are termed as `subohmic', `ohmic' and
`superohmic' baths respectively. We will focus on ohmic bath as an
example given in Sec. \ref{sec:example}.

To derive the master equation, we first consider the initial total
density matrix of the factorizable form:
\begin{eqnarray}
\rho_{tot}(0) & = & | \alpha_0 \rangle \langle \alpha_0| \nonumber \\ & & \otimes
\prod_{k} \left[ (e^{-\hbar \omega_{k} b_{k}^{\dagger} b_{k} / k_{B}
T })(1-e^{-\hbar \omega_{k}/ k_{B} T }) \right]
\label{eq:Gaussianthermalbath}
\end{eqnarray}
where $| \alpha_0 \rangle$ is a coherent state for system and the
bath is initially in thermal equilibrium at temperature $T$. At
later time $t > 0$, the total density matrix in position-space
representation reads
\begin{eqnarray}
& & \rho_{tot} (X'', {\bf q''}, X', {\bf q'}, t) \nonumber \\ & = & \int \! dX d{\bf
q} \int \! dX''' d{\bf q'''} \, U(X'',{\bf q''},t;X,{\bf q},0)
\nonumber \\ & & \times \rho_{tot} (X,{\bf q}, X''', {\bf q'''}, 0)
U^{\ast}(X',{\bf q'},t;X''',{\bf q'''},0) \label{eq:rho_tot}
\end{eqnarray}
where $\rho_{tot} (X'', {\bf q''}, X', {\bf q'}, 0) \equiv \langle
X'',{\bf q''} | \rho_{tot}(0) | X', {\bf q'} \rangle$, where ${\bf
q} \equiv (\{ q_{k} \})$ and $U(X'',{\bf q''},t;X',{\bf q'},0)
\equiv \langle X'',{\bf q''} | {\cal T}
 \exp \left[ -i \int_{0}^{t} \! ds \, H(s) \right] |
X', {\bf q'} \rangle$, with ${\cal T}$ being the time-ordering
operator. The factorized initial density matrix
(\ref{eq:Gaussianthermalbath}) guarantees that the Liouville
operator is independent of initial system state, which was also
observed in \cite{HPZ, Connell, Yu}. The Guassian initial state
(\ref{eq:Gaussianthermalbath}) and the Gaussian kernel in Eq.
(\ref{eq:rho_tot}) resulted from the linearity of total Hamiltonian
allow exact integration, which makes the reduced density matrix
$\rho_{S} = {\rm Tr}_{B} (\rho_{tot})$ also a Gaussian. In other
words, the master equation governing the evolution of $\rho_{S}$
must preserve the Gaussian properties. This requires that the master
equation involves only some quadratic combinations of $a$ and
$a^{\dagger}$ as in the HPZ master equations.  Together with the
requirements of conservation of probability [${\rm Tr} (
\dot{\rho}_{S} ) = 0$], hermiticity ($\rho_{S} =
\rho_{S}^{\dagger}$) and state-independent coefficients
($\gamma_{1}$, $\gamma_{2}$, $\gamma_{3}$, $\Delta \omega$ and
$\Delta \lambda$ in below), we have the following form of
time-convolutionless master equation:
\begin{eqnarray}
\dot \rho_{S} & = & - i \left[ {H_{S} (t)  + \Delta H_{S} (t)
,\rho_{S} } \right] \nonumber \\
& & - \gamma_{1} (t) \left( {a^{\dagger}  a\rho_{S}  + \rho_{S}
a^{\dagger} a - 2a\rho_{S} a^{\dagger}  } \right) \nonumber \\
& & - \gamma_{2} (t) \left( {aa^{\dagger}
\rho_{S}  + \rho_{S} aa^{\dagger}   - 2a^{\dagger}  \rho_{S} a} \right) \nonumber \\
& & - \gamma_{3} (t) \left( {aa\rho_{S}  + \rho_{S} aa - 2a\rho_{S}
a} \right) \nonumber \\
& & - \gamma_{3}^{\ast} (t) \left( {a^{\dagger}  a^{\dagger}
\rho_{S}  + \rho_{S} a^{\dagger}  a^{\dagger}   - 2a^{\dagger}
\rho_{S} a^{\dagger}  } \right) \label{eq:mastereq}
\end{eqnarray}
where
\begin{equation}
\Delta H_{S} (t)  = \frac{{\Delta \lambda (t)}}{2}a^{2}  +
\frac{{\Delta \lambda ^* (t) }}{2}a^{\dagger 2}  + \Delta \omega (t)
a^{\dagger} a \label{eq:deltaHS}
\end{equation}
are modifications to the system Hamiltonian due to system-bath
interaction. In particular, $\Delta \omega (t)$ is the frequency
shift term, $\Delta \lambda (t)$ modifies the parametric
interaction, $\gamma_{1}(t)$ and  $\gamma_{2} (t)$ are functions
governing dissipation and amplification, and the terms with
$\gamma_{3}(t)$ describe phase-dependent decoherence typically
appear in squeezed baths \cite{Gardiner}.

Our next task is to determine the time-dependent coefficients. To
this end, with the formal solution of $b_{k}(t)$, let us write down
the Heisenberg's equation for $a(t)$,
\begin{eqnarray}
\dot{a} (t) & = & - i \lambda^{\ast} (t) a^{\dagger} (t) \! - \!i \left[ \omega_{0}
+ \Omega (t) \right] a (t) \nonumber \\ & & - i \sum_{k} g_{k} \left[ b_{k}(0) e^{-i \omega_{k} t}
+ b_{k}^{\dagger}(0) e^{i \omega_{k} t} \right] \nonumber \\
\! & & - \! \int_{0}^{t} \! ds \, K (t-s) \left[ a(s)+a^{\dagger}(s)\right]
\label{eq:withoutRWA}
\end{eqnarray}
where $K (\tau)$ is the memory kernel and $K (\tau) \equiv -2i \sum_{k}
g_{k}^{2} \sin \omega_{k} \tau = -2i\int_{0}^{\infty} \! d\omega \, J(\omega)
\sin \omega \tau $.

The linearity of Eq. (\ref{eq:withoutRWA}) leads to a general operator solution for
$a(t \geq 0)$ that can be expressed in terms of initial conditions,
\begin{equation}
{a}(t) = G(t)\, {a}(0) + L^{\ast}\!(t)\, a^{\dagger}(0) + {F}(t)
\label{eq:at}
\end{equation}
where $F(t) = \sum_{k} \left( \mu_{k} (t) b_{k} (0) + \nu_{k}^{\ast}
(t) b_{k}^{\dagger} (0) \right)$. The functions $G(t)$, $L(t)$,
$\mu_{k}(t)$ and $\nu_{k} (t)$ can be determined by substituting Eq.
(\ref{eq:at}) into Eq. (\ref{eq:withoutRWA}) and comparing
coefficients of initial system operators. This gives,
\begin{eqnarray}
\dot{G} & = & -i \lambda^{\ast} (t) L -i \left[ \omega_{0} + \Omega (t) \right]
G \nonumber \\
& & - \! \int_{0}^{t} \! ds \, K (t-s) \, \left[ G(s)+L(s) \right], \label{eq:Gt} \\
\dot{L} & = & i \lambda (t) G +i \left[ \omega_{0} + \Omega (t) \right] L \nonumber \\
& & + \!
\int_{0}^{t} \! ds \, K (t-s) \, \left[ G(s)+L(s) \right], \label{eq:Lt} \\
\dot{F} & = & -i \lambda^{\ast} (t) F^{\dagger} -i \left[
\omega_{0} + \Omega (t) \right] F \nonumber \\
& & - \! \int_{0}^{t} \! ds \, K (t-s)
\left[ F(s) + F^{\dagger}(s) \right] \nonumber \\
& & - i \sum_{k} g_{k} \left[ b_{k}(0) e^{-i \omega_{k} t} +
b^{\dagger}_{k}(0) e^{i \omega_{k} t} \right] \label{eq:Ft}
\end{eqnarray}
with initial conditions $G(0)=1$, $L(0)=0$, $\mu_{k}(0)=0$ and
$\nu_{k} (0)=0$ (see Appendix \ref{sec:GLF}
for further reduction on $F(t)$). Hence for a given spectral density $J(\omega)$ and
system Hamiltonian, $G(t)$ and $L(t)$ can be solved and the operator
solution $a(t)$ can be found.

Now we make use of the fact that the time derivative for $\langle
a(t) \rangle$, $\langle a(t)a(t) \rangle$ and $\langle
a^{\dagger}(t)a(t) \rangle$ obtained by the master equation
(\ref{eq:mastereq}) must agree with that obtained by the
corresponding Heisenberg operator solution after taking expectation
values. A direct comparison of these equations (see Appendix
\ref{sec:coeffcom} for details) allows us to determine $\gamma_{1}$,
$\gamma_{2}$, $\gamma_{3}$, $\Delta \omega$ and $\Delta \lambda$. It
is easy to show in the comparison that
\begin{equation}
\Delta \lambda = \Delta \omega -i (\gamma_{1} - \gamma_{2})
\end{equation}
which is not independent from other time-dependent coefficients, and the
remaining coefficients are:
\begin{eqnarray}
2i\Delta\omega &=& \frac{1}{W(t)}\int_{0}^{t}\! ds \, K (t-s) \, \nonumber \\ & &
\{\left[G(t)-L(t)\right]\left[L^{\ast}(s)+G^{\ast}(s)\right] \nonumber \\ & & \hspace{5mm} +\left[G^{\ast}(t)-L^{\ast}(t)\right]\left[G(s)+L(s)\right]\}  \label{eq:Domega}  \\
\gamma_{1} &=& -\frac{1}{2}\frac{\dot{W}(t)}{W(t)}+\gamma_{2} \\
2\gamma_{2} & = & \frac{d}{dt}\langle F^{\dagger}F \rangle-\frac{\dot{W}(t)}{W(t)}
\langle F^{\dagger}F \rangle \nonumber \\
& & -\frac{\dot{L}G^{\ast}-\dot{G}^{\ast}L}{W(t)}\langle
FF \rangle \nonumber \\ & & - \frac{\dot{L}^{\ast}G-\dot{G}L^{\ast}}{W(t)}\langle F^{\dagger}F^{\dagger}
\rangle \\
-2\gamma_{3}^{\ast} & = & \frac{d}{dt}\langle FF \rangle -2\frac{\dot{G}G^{\ast}-
\dot{L}^{\ast}L}{W(t)}\langle FF \rangle \nonumber \\
& & -\frac{\dot{L}^{\ast}G-\dot{G}L^{\ast}}{W(t)}
\langle F^{\dagger}F+FF^{\dagger} \rangle \label{eq:gamma3}
\end{eqnarray}
where $W(t) \equiv G(t) G^{\ast} (t) - L(t) L^{\ast} (t)$ and the bath-bath
correlation functions are
\begin{eqnarray}
\langle F(t) F(t) \rangle & = & - \int_{0}^{t}\! ds' \int_{0}^{t}\! ds'' \,
\kappa_{T} (s''-s') \nonumber \\
& & \left[ G(s') - L^{\ast} (s') \right] \left[ G(s'') -
L^{\ast} (s'') \right] \\
\langle F^{\dagger}(t) F(t) \rangle & = & \int_{0}^{t}\! ds' \int_{0}^{t}\! ds''
\, \kappa_{T} (s''-s') \nonumber \\
& & \left[ G^{\ast}(s') - L (s') \right] \left[ G(s'') -
L^{\ast} (s'') \right] \\
\langle F(t) F^{\dagger}(t) \rangle & = & \int_{0}^{t}\! ds' \int_{0}^{t}\! ds''
\, \kappa_{T} (s''-s') \nonumber \\
& & \left[ G(s') - L^{\ast} (s') \right] \left[ G^{\ast}(s'')
- L (s'') \right]
\end{eqnarray}
The temperature-dependent memory kernel $\kappa_{T}(\tau)$ takes the form
\begin{equation}
\kappa_{T} (\tau) \equiv \sum_{k} g_{k}^{2} \left[ 2 \cos \left( \omega_{k}
\tau \right) \left( e^{\frac{\hbar \omega_{k}}{k_{B}T}}-1 \right)^{-1} +
e^{-i \omega_{k} \tau } \right]
\end{equation}
for a thermal bath at temperature $T$. In particular, at zero temperature,
\begin{equation}
\kappa_{0} (\tau) \equiv \sum_{k} g_{k}^{2} e^{-i \omega_{k} \tau }.
\end{equation}
In the special case $\Omega (t)= \lambda (t)=0$, we have ${\rm Re}
(\gamma_{3}) = (\gamma_{1}+\gamma_{2})/2$ and our equation can be
reduced to HPZ master equation.

Since the master equation (\ref{eq:mastereq}) works for any initial
system coherent state $|\alpha_0 \rangle$, we can generalize the
results to arbitrary initial system states by using the
Glauber-Sudarshan P-representation. This is because for an arbitrary
system state, we can formally express its density matrix in the
diagonal form,
\begin{equation}
\rho_{S} (0) = \int \! d^{2} \alpha \, P(\alpha) |\alpha \rangle
\langle \alpha |.
\end{equation}
By the linearity of (\ref{eq:mastereq}) and the fact that all the
coefficients are independent of initial $\alpha_0$, we can conclude
that the master equation (\ref{eq:mastereq}) is also valid for any
initial system states.

\section{Example: Parity kick control}

\label{sec:example}

\begin{figure}[h]
\centering
\includegraphics[width=6 cm]{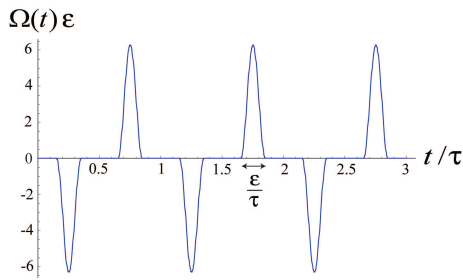}
\caption{Sketch of the pulses in three periods with $\tau /2$ being the
time between successive peaks. Here $\epsilon / \tau = 0.2$ is
used.} \label{fig:omegat}
\end{figure}

The master equation (\ref{eq:mastereq}) with the coefficients given
in Eq. (\ref{eq:Domega}-\ref{eq:gamma3}) is the main result of this paper. Such an equation
provides a useful tool to determine the behavior of a damped
harmonic oscillator subjected to external modulation of system
parameters. To provide an explicit example, we employ our master
equation to study the dynamics in `parity kick' decoherence control
problems. Previous studies of this subject were mostly confined to
ideal $\delta$-pulses or square pulses that have finite
discontinuous jumps \cite{Tombesi1, Tombesi2, ShiokawaHu}. In this
section we examine parity kick with soft pulses and its efficiency.
Specifically, we consider the system Hamiltonian $H_{S} (t)$ in Eq.
(\ref{eq:HS}) with $\lambda (t) = 0$ and
\begin{equation}
\Omega (t) = \sum_{n} \frac{2 \pi}{\epsilon} \{ - \phi( \epsilon, n \tau+\frac{\tau}{4},
t) + \phi( \epsilon, n \tau+\frac{3\tau}{4}, t) \}
\end{equation}
where
\begin{eqnarray}
\phi(\epsilon, s, t) & \equiv & \theta \left[ t- \left( s -
\frac{\epsilon}{2} \right) \right] \theta \left[\left( s + \frac{\epsilon}{2}
\right) - t \right] \nonumber \\ & & \times \sin ^{2} \left[ \frac{\pi}
{\epsilon} \left( t- s + \frac{\epsilon}{2} \right)
\right]
\end{eqnarray}
and $\theta (t)$ is the unit-step function. The driving frequency
$\Omega (t)$ corresponds to a pair of sine-squared pulses within one
pulse period $n \tau \leq t < (n+1) \tau$, each with pulse duration
characterized by $\epsilon$, and they peak at $t=n \tau
+\frac{\tau}{4}$ and $t=n\tau +\frac{3\tau}{4}$ with strengths
$-\pi$ and $\pi$ respectively (see Fig. \ref{fig:omegat}). The idea
of using pulses with alternating signs has been discussed in
\cite{ShiokawaHu}. The pair of $\pi$ pulses with zero pulse width
($\epsilon \rightarrow 0$) can effectively reverse the direction of
interaction. If the system and bath start uncoupled, then one can
prevent the system from coupling with the bath at later times by
applying Dirac-delta shaped pulses frequently.

For ideal Dirac-delta shaped pulses $(\epsilon \to 0)$, the pulse
width is so short and the strength so strong as that Eqs.
(\ref{eq:Gt}) and (\ref{eq:Lt}) yield $\dot{G}(t) \approx - i \pi
\delta (t-t_{0}) G(t)$  and $\dot{L}(t) \approx i \pi \delta
(t-t_{0}) L(t)$ around $t=t_{0}$ where the $\pi$-strength pulse
peaks. Integrating, we observe that the sole effect of such an ideal
pulse is to flip the sign of $G(t)$ and $L(t)$, while leaving
$\dot{G}(t)$ and $\dot{L}(t)$ unchanged, i.e., $G(t_{0}^{-})
\rightarrow -G(t_{0}^{+})$, $\dot{G} (t_{0}^{-}) \rightarrow
\dot{G}(t_{0}^{+})$, $L(t_{0}^{-}) \rightarrow -L(t_{0}^{+})$ and $
\dot{L}(t_{0}^{-}) \rightarrow \dot{L}(t_{0}^{+})$. This leads to
the sign-flip of coefficients $\gamma_{1}$, $\gamma_{2}$,
$\gamma_{3}$ and $\Delta \omega$ of the above master equation.

However, we point out that while the coefficients of master equation
(\ref{eq:mastereq}) acquire negative sign immediately after the
kick, it does not necessarily constitute an effective scheme in
coherence protection. Typically memory time characterizes the
transient time before the coefficients eventually settle at their
long-time limits with the system decaying steadily. Frequent kicking
leads to quick and repeating sign-flips that inhibits the free
evolution of the coefficients $\gamma_j(t)$ and their subsequent
settlements, which also produces sawtooth-like graphs for the
coefficients. The dissipative coefficients would average to zero
over an extended period of time and thus protect the system state
and coherence from decaying. Specifically, for an effective scheme
the kicking period $\tau$ should be less than the memory time, which
is of the order $1/\omega_{c}$, in order to achieve good coherence
protection. Vitali and Tombesi \cite{Tombesi1} also pointed out that
high-frequency bath oscillators with $\omega_{k} \gg \omega_{0}$
would not be able to evolve significantly before interaction
Hamiltonian changes sign if a kicking frequency much higher than
cut-off frequency is used. Memory time, or equivalently the inverse
of cut-off frequency, then plays an important role in fixing the
coherence protection scheme. In light of this, the ideal decoupling
pulse sequence would consist of pulses with width $\epsilon
\rightarrow 0$ and pulse period $\tau \ll 1/\omega_{c}$. More
discussions about the use of $\delta$-pulses can be found in
\cite{Tombesi1, Tombesi2, ShiokawaHu}.

\begin{figure}[h]
\centering
\includegraphics[width=6cm]{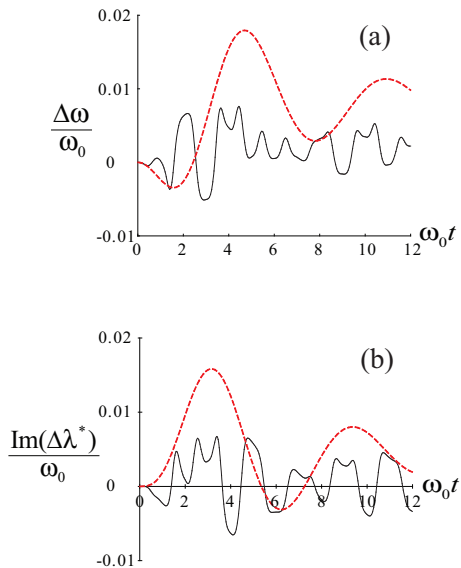}
\caption{(Colour online) Black solid line: Time-dependence of
modification of oscillation frequency (a) and the imaginary part of
parametric coupling (b), with $\eta = 0.2$, $\omega_{c} / \omega_{0}
= 0.2$. $\epsilon / \tau = 0.5$ and $\omega_{0} \tau = 2.0$. Red
dashed line correspond to cases without parity kicks for comparison.
The bath is ohmic at zero temperature.} \label{fig:wide5A}
\end{figure}

\begin{figure}[h]
\centering
\includegraphics[width=8cm]{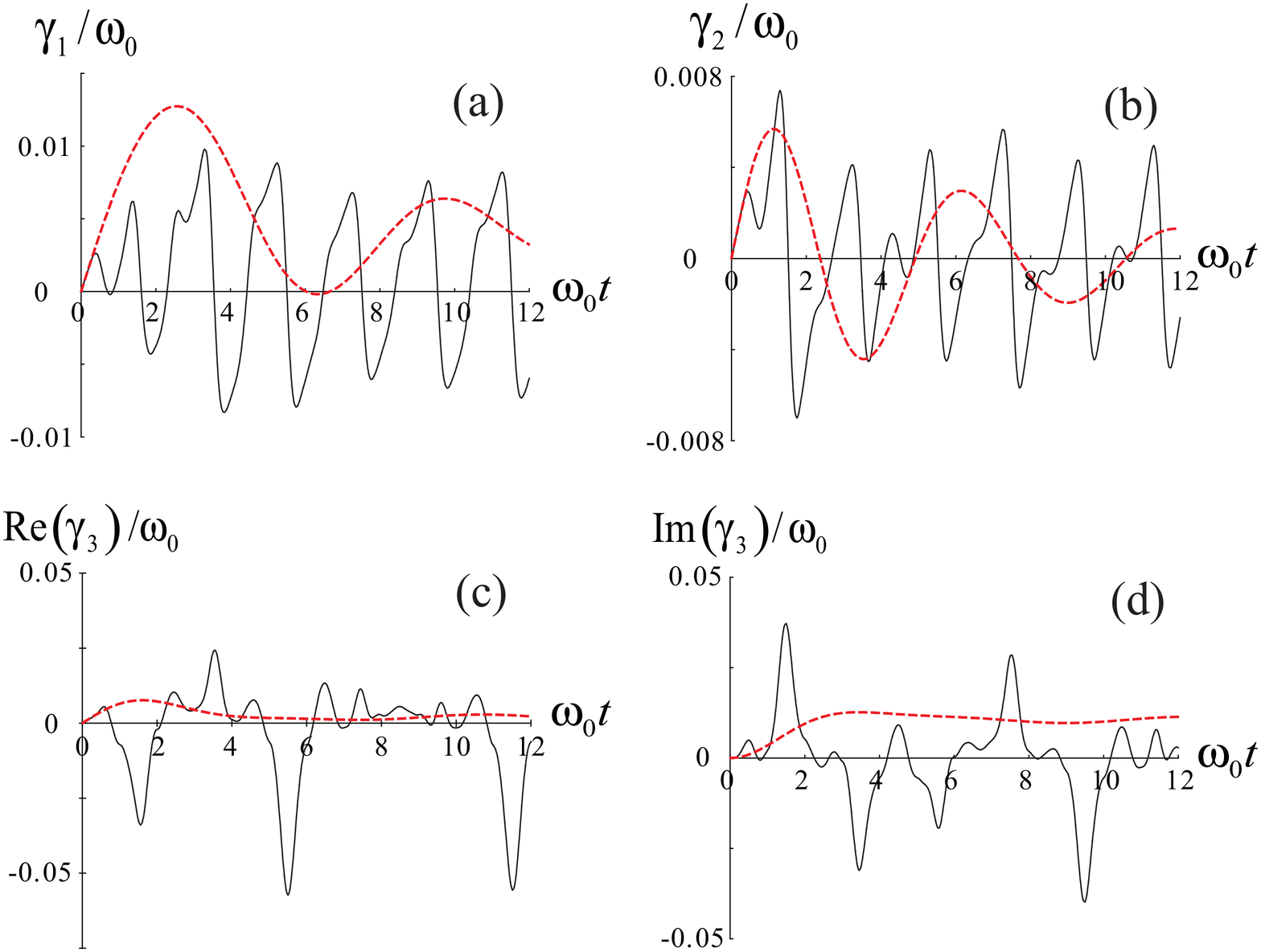}
\caption{(Colour online) Black solid line: Time-dependence of
relaxation coefficients of the master equation for an ohmic bath at
zero temperature with $\eta = 0.2$, $\omega_{c} / \omega_{0} = 0.2$,
$\epsilon / \tau = 0.5$ and $\omega_{0} \tau = 2.0$. Red dashed
lines correspond to cases without parity kicks for comparison. The
bath is ohmic at zero temperature.} \label{fig:wide5B}
\end{figure}

For soft pulses, we can obtain numerically the time-dependent
coefficients of master equation according to the prescription in
Sec. \ref{sec:model}. An example is given in Figs. \ref{fig:wide5A}
and \ref{fig:wide5B} in which we can compare the coefficients of
master equation under the influence of soft-pulse kicks with the
free-evolution scenario. In our calculations, we consider an ohmic
bath at zero temperature. The parameters $\eta$ and $\omega_{c}$ are
chosen such that non-Markovian features, including transient
behaviour and non-exponential decay of coherence, can be observed
for several natural oscillation cycles. In Fig. \ref{fig:wide5A} we
show the modifications of system Hamiltonian $\Delta H_S$ in Eq. (\ref{eq:deltaHS})
due to the bath interaction. Without kicking (red dashed lines),
$\Delta \omega$ and $\Delta \lambda$ are seen to exhibit oscillatory
patterns with frequency close to the natural frequency $\omega_{0}$
of the system. The effect of parity kicks (black solid lines) seem
to suppress partially both $\Delta \omega$ and $\Delta \lambda$ with
complicated oscillations that follow the kicking frequency $1/ \tau$
when `parity kick' is in place. In Fig. \ref{fig:wide5B} the
time-dependence of dissipative coefficients $\gamma_1$, $\gamma_2$
and $\gamma_3$ are shown. We note that perfect sign-flip is not
present in the figures due to the use of soft pulses. Instead,
$\gamma_{1}$ and $\gamma_{2}$ display sawtooth-like periodic
oscillations, which average to numbers close to zero in the long
run.

\begin{figure}[h]
\centering
\includegraphics[width=7 cm]{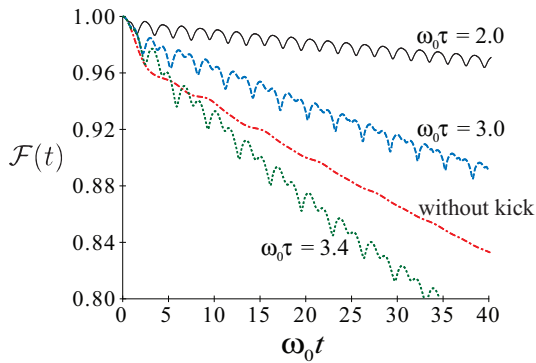}
\caption{(Color online)  Evolution of fidelity ${\cal F} (t)$
subjected to soft pulses with $\omega_{0} \epsilon = 1.0$ and
various kicking period $\tau$. The parameters are: $\eta = 0.2$,
$\omega_{c} / \omega_{0} = 0.2$ and an ohmic bath at zero
temperature.} \label{fig:widefid}
\end{figure}

With the coefficients solved, we proceed to examine the dissipation
decay of an initially excited system. As an illustration, we
consider the initial state $|\psi_{S} (0) \rangle = |1\rangle $,
where $|1\rangle$ is the first excited energy eigenstate of the
system. The efficiency of decoherence protection is quantified by
the fidelity ${\cal F} (t) \equiv \sqrt{\langle \psi_{S} (0) |
\rho_{S}(t) | \psi_{S} (0) \rangle} $ for the pure initial system
state. For an ohmic bath and at zero temperature, we observe in Fig.
{\ref{fig:widefid}} that free evolution (without kicks) no longer
gives an exponentially decaying fidelity, and the initial slip is
obvious. Our numerical results indicate that parity kicks with soft
pulses can suppress the decay if $\tau$ is sufficiently short.
However, we notice a transition from coherence protection to
decoherence acceleration with increasing pulse period, despite the
fact that the pulse periods used in Fig. \ref{fig:widefid} are all
shorter than characteristic memory time of the bath. This is related
to the anti-Zeno effect \cite{antizeno}, and it has been discussed
the $\delta$-pulse limit in Ref. \cite{ShiokawaHu} where the authors
suggested that this transition, which occurs when $\tau \sim 1/
\omega_{c}$, is not unique to ohmic bath. We have also performed
other simulations (not shown) for soft pulses with various widths.
For the same parameters in Fig. \ref{fig:widefid}, we find that
increasing the width $\epsilon$ would slightly decrease the
fidelity.

\section{Conclusion}

We have derived an exact non-Markovian master equation for a damped
harmonic oscillator linearly coupled with an oscillator thermal
bath, with a time-dependent natural frequency and parametric
interaction strength. The equation enables us to determine the
evolution of system density matrix  under the influence of various
pulses or signals. Expressions of the coefficients of master
equation as well as the correlation functions are computable once we
solve the functions $G(t)$ and $L(t)$. We emphasize that the method
works because of the linearity of the whole system, which leads to
the Gaussian propagator, so that we need only to consider a finite
number of equations. Our approach has been applied to parity kick
problems using soft pulses, and we observe that coefficients of
master equations exhibit complicated time-dependence when the system
frequency is modulated by $\pi$-pulses. In particular, our numerical
results suggest that suppression of decoherence can be achieved by
soft pulses, as long as the kicking frequency sufficiently higher
than cut-off frequency of the bath.

\begin{acknowledgments}
The work described in this paper was supported by a grant from the
Research Grants Council of Hong Kong, Special Administrative Region
of China (Project No. CUHK401307).
\end{acknowledgments}

\appendix

\section{derivation of $G$, $L$ and $F$}
\label{sec:GLF}
With the Heisenberg's equation of motion of system operator $a(t)$, by
combining (\ref{eq:withoutRWA}) and (\ref{eq:at}) we can write down equations (\ref{eq:Gt}-\ref{eq:Ft})
by comparing coefficients of operators $a(0)$ and
$a^{\dagger}(0)$. From the expectation value of (\ref{eq:at}), we have the initial conditions
$G(0)=1$, $L(0)=0$ and $\langle F(0) \rangle =0$, which would solve the
differential equations for $G(t)$ and $L(t)$ above given the memory kernel
$K(\tau)$. The bath operator $F(t)$ is solved by Green's function approach.
We let
\begin{equation}
{\cal \vec{V}} = \int_{0}^{t}\! dt'\, {\bf \Gamma} (t-t') \,
{\cal \vec{B}}(t')  \label{eq:Ft_vec}
\end{equation}
where
\begin{equation}
{\cal \vec{V}} \equiv \left( F, F^{\dagger} \right)^{T}
\end{equation}
and
\begin{eqnarray}
{\cal \vec{B}}(t) & \equiv & \left( B(t), -B^{\dagger}(t) \right)^{T}
\nonumber \\
& \equiv & \sum_{k} g_{k} \left[ e^{-i \omega_{k} t} b_{k}(0) + e^{i
\omega_{k} t} b_{k}^{\dagger}(0) \right] \begin{bmatrix}1\\
                           -1\\ \end{bmatrix}
\end{eqnarray}
The $2 \times 2$ matrix $\bf \Gamma (\tau)$ has two independent entries, which may
be viewed as the Green functions. Differentiating (\ref{eq:Ft_vec}) w.r.t.
$t$ and setting ${\bf \Gamma}(0)= -i {\bf I}$, where ${\bf I}$ is the $2 \times 2$ identity matrix, we arrive at
\begin{eqnarray}
& & \int_{0}^{t} [i\dot{{\bf \Gamma}} (t-t') - {\bf M}{\bf \Gamma}(t-t')
\nonumber \\ & & + i \! \int_{0}^{t} \! ds \, {\bf K}(t-s) \, {\bf \Gamma}(s-t') ] {\cal \vec{B}}(t') dt' = 0
\label{eq:GF}
\end{eqnarray}
with the matrices ${\bf M} (\tau)$ and ${\bf K}(\tau)$ defined as
\begin{equation}
{\bf M} (\tau) = \begin{bmatrix}\omega_{0} + \Omega (\tau) &\lambda^{\ast}
(\tau)\\
                           -\lambda (\tau) &-\omega_{0} - \Omega (\tau)\\
                           \end{bmatrix}
\end{equation}
\begin{equation}
{\bf K}(\tau) = K (t-s) \begin{bmatrix}1&1\\
                           -1&-1\\ \end{bmatrix}
\end{equation}
Since the L.H.S. of equation (\ref{eq:GF}) must vanish for all ${\cal
\vec{B}}(\tau)$, we have
\begin{equation}
i\dot{{\bf \Gamma}} (t-t') - {\bf M}{\bf \Gamma}(t-t') + i \! \int_{0}^{t}
\! ds \, {\bf K}(t-s) \, {\bf \Gamma}(s-t') = 0
\end{equation}
While one can solve the Green's functions with the initial condition stated
above, a simple comparison with the differential equations for $G(t)$ and
$L(t)$ immediately yields the following correspondence
\begin{equation}
{\bf \Gamma}(\tau) = -i \begin{bmatrix} G(\tau) & L^{\ast}(\tau)\\
                           L(\tau) & G^{\ast}(\tau) \\ \end{bmatrix}
\label{eq:GammatoGL}
\end{equation}
Thus the bath operator $F(t)$ can be completely solved if coupling constants
are given and $G(\tau)$ and $L(\tau)$ are known. Explicitly,
\begin{eqnarray}
F(t) = &- &i \int_{0}^{t} ds \left[ G(t-s)-L^{\ast}(t-s) \right] \nonumber \\ & \times & \sum_{k}
g_{k} \left[ b_{k}(0)e^{-i \omega_{k}s} + b_{k}^{\dagger}(0)e^{i \omega_{k}s}
\right]
\end{eqnarray}

\section{comparison of coefficients}
\label{sec:coeffcom}
From the operator solution of $a(t)$ (\ref{eq:at}), taking derivatives
w.r.t. $t$ and eliminating operators at $t=0$, we have the following
operator equation
\begin{eqnarray}
\dot{a} = & &\frac{\dot{G}G^{\ast}-\dot{L}^{\ast}L}{W(t)}
a + \frac{\dot{L}^{\ast}G-\dot{G}L^{\ast}}{W(t)}
a^{\dagger} \nonumber \\ & & + \dot{F}-\frac{\dot{G}G^{\ast}-\dot{L}^{\ast}L}
{W(t)} F -\frac{\dot{L}^{\ast}G-\dot{G}L^{\ast}}
{W(t)} F^{\dagger}
\label{eq:Hdat}
\end{eqnarray}
where $W(t) \equiv G(t) G^{\ast} (t) - L(t) L^{\ast} (t)$. This operator
equation also enables us to express second moments' equations in terms of
$G(t)$ and $L(t)$. Expectation values are taken with respect to the
initially factorizable state, with the bath at thermal equilibrium.
With this choice of initial state, system-bath correlations would be
reduced to bath-bath correlations, e.g.
\begin{equation}
\langle a(t)F(t) \rangle = \langle F(t)F(t) \rangle.
\label{eq:sbtobb}
\end{equation}
This follows from the fact that ${\rm{Tr_{B}}}[\rho_{B}(0)F(t)]$ vanishes,
and (\ref{eq:sbtobb}) is an essential condition that guarantees a
state-independent master equation. We proceed to write down the equations
obtained from the master equation (\ref{eq:mastereq}) and equation
(\ref{eq:Hdat}).
From the master equation (\ref{eq:mastereq}),
\begin{eqnarray}
\frac{d}{dt}\langle a \rangle & = & -[\xi (t)+i\chi (t)] \langle a \rangle - i\Lambda^{\ast} (t)\langle a^{\dagger} \rangle \\
\frac{d}{dt}\langle aa \rangle &=& -2[\xi (t) +i\chi (t)] \langle
aa \rangle -2i\Lambda^{\ast} (t)\langle a^{\dagger}a \rangle \nonumber \\ & & -
i\Lambda^{\ast} (t)-2\gamma_{3}^{\ast} (t) \\
\frac{d}{dt}\langle a^{\dagger}a \rangle &=& -2\xi (t) \langle a^{\dagger}a \rangle + i\Lambda (t)
\langle aa \rangle \nonumber \\ & & - i\Lambda^{\ast} (t) \langle a^{\dagger}a^{\dagger} \rangle + 2\gamma_{2} (t)
\end{eqnarray}
where $\xi (t) \equiv \gamma_{1} (t) - \gamma_{2} (t)$, $\chi (t) \equiv \omega_{0}
+ \Omega (t) + \Delta \omega (t)$ and $\Lambda (t) \equiv \lambda (t) + \Delta \lambda (t)$.
and from the Heisenberg's equation (\ref{eq:Hdat}),
\begin{eqnarray}
\frac{d}{dt}\langle a \rangle & = &
\frac{\dot{G}G^{\ast}-\dot{L}^{\ast}L} {W(t)} \langle a \rangle +
\frac{\dot{L}^{\ast}G-\dot{G}L^{\ast}}
{W(t)} \langle a^{\dagger} \rangle \\
\frac{d}{dt}\langle aa \rangle & = & 2\frac{\dot{G}G^{\ast}-\dot{L}^{\ast}L}
{W(t)}\langle aa \rangle \nonumber \\ & &+ 2\frac{\dot{L}^{\ast}G-\dot{G}L^{\ast}}
{W(t)}\langle a^{\dagger}a \rangle+\frac{\dot{L}^{\ast}G-
\dot{G}L^{\ast}}{W(t)} \nonumber \\ & & +  \frac{d}{dt}
\langle FF \rangle -2\frac{\dot{G}G^{\ast}-\dot{L}^{\ast}L}{W(t)}\langle FF
\rangle \nonumber \\ & &-\frac{\dot{L}^{\ast}G-\dot{G}L^{\ast}}
{W(t)}\langle F^{\dagger}F+FF^{\dagger} \rangle \\
\frac{d}{dt}\langle a^{\dagger}a \rangle & = & \frac{\dot{W}(t)}{W(t)}
\langle a^{\dagger}a \rangle +\frac{\dot{L}
G^{\ast}-\dot{G}^{\ast}L}{W(t)}\langle aa \rangle \nonumber \\ & &+
\frac{\dot{L}^{\ast}G-\dot{G}L^{\ast}}{W(t)}\langle
a^{\dagger}a^{\dagger} \rangle + \frac{d}{dt}\langle
F^{\dagger}F \rangle \nonumber \\ & &-\frac{\dot{W}(t)}{W(t)}
\langle F^{\dagger}F \rangle-\frac{\dot{L}G^{\ast}-\dot{G}^{\ast}L}
{W(t)}\langle FF \rangle \nonumber \\ & & - \frac{\dot{L}^
{\ast}G-\dot{G}L^{\ast}}{W(t)}\langle F^{\dagger}F^{\dagger}
\rangle
\end{eqnarray}
where $G$ and $L$ are understood to be time-dependent functions. By comparing
these two sets of equations we obtain the coefficients of master equation
(\ref{eq:mastereq}).

\end{document}